\documentclass[twocolumn,showpacs,preprintnumbers,prl]{revtex4}
\usepackage{graphicx}
\begin{document}
\title{Current-induced torques due to compensated antiferromagnets}
\author{Paul M. Haney and  A. H. MacDonald}
\address{ Department of Physics, The University of Texas at Austin, Austin,
Texas, 78712-0264, U.S.A. }

\begin{abstract}
We analyse the influence of current induced torques on the
magnetization configuration of a ferromagnet in a circuit
containing a compensated antiferromagnet.  We argue that these
torques are generically non-zero and support this conclusion with
a microscopic NEGF calculation for a circuit containing
antiferromagnetic NiMn and ferromagnetic Co layers. Because of
symmetry dictated differences in the form of the current-induced
torque, the phase diagram which expresses the dependence of
ferromagnet configuration on current and external magnetic field
differs qualitatively from its ferromagnet-only counterpart.
\end{abstract}

\maketitle

\noindent {\em Introduction}--- Current-induced torques in
noncollinear ferromagnetic metal circuits were predicted over 10
years ago \cite{slonc,berger}, and have since been the subject of
an extensive and quite successful body of experimental and
theoretical research.  Almost all studies of current-induced
torques consider either their role in ferromagnetic (F) spin valve
circuits \cite{tsoi1,tsoi2,katine,albert,pufall} or their
influence on magnetic domain wall motion \cite{beach, grollier,
yamaguchi, yamanouchi,hayashi}. In both cases, the current-induced
torques can be understood as following from the transfer of
conserved spin angular momentum from current-carrying
quasiparticles to the magnetic condensate, hence the term
spin-transfer torque. It has recently been predicted that
current-induced torques are generically present whenever
non-equilibrium quasiparticles interact with non-collinear
magnetic order parameters, even\cite{alvaroAFM,haneyAFM} in
circuits containing only antiferromagnetic (AF) elements.
Experiments\cite{tsoiAFM} have established a dependence of
unidirectional exchange bias fields on current, providing indirect
evidence that current-induced torques are present in AFs. In this
Letter we analyse the influence of current-induced torques on a F
thin film in a circuit containing a compensated AF. Because of a
key difference in symmetry compared to purely F spin-valve
circuits, we find that the phase diagram which expresses the
dependence of the magnetic configuration of the F on current and
external magnetic field differs qualitatively from the familiar F
only spin-valve phase diagram\cite{kiselev}. In particular, we
find that transport currents can drive the F to a stable steady
state with magnetization perpendicular to the AF layer moments. In
the following paragraphs we argue on symmetry grounds for the form
of the current-induced torque, and explore its robustness by
performing a fully microscopic current-induced torque calculation
for a circuit containing Co and NiMn layers. We then turn our
attention to the construction of the F state phase diagram implied
by equations of motion which include the current-induced torque
term, and conclude with a discussion of experimental implications.

\begin{figure}[h!]
\begin{center}
\vskip 0.2 cm
\includegraphics[width=2.5in]{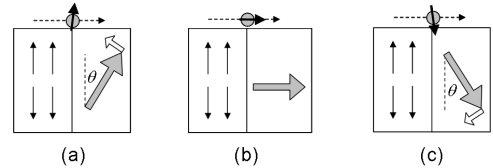}
\vskip 0.2 cm \caption{Current-induced torques due to a
compensated antiferromagnet.  The arrows above the structure
indicate the electron flux spin direction.  The white arrows
indicate the ensuing current-induced torques on the FM.}
\label{fig:sttcartoon}
\end{center}
\end{figure}

\noindent {\em Current-induced Torques due to Compensated
Antiferromagnets}--- The total current-induced torque acting on a
F nanoparticle can always\cite{slonc} be expressed in terms of the
difference between incoming and outgoing spin currents. The
presence of a ferromagnet will in general induce a nonzero spin
current at the AF-F interface.  When spin-polarized electron flux
from an AF with orientation ${\hat n}_{AF}$ enters a F with
orientation ${\hat n}_{F}$, the spin current entering F will have
some component in the ${\hat n}_{AF}$ direction.  It follows that,
just as in the familiar case where both materials are F, a
current-induced torque will act in the plane defined by ${\hat
n}_{AF}$ and ${\hat n}_{F}$, as illustrated in Fig.
(\ref{fig:sttcartoon}). (Out of plane torques are also non-zero
but tend to be much smaller.) Spin-invariance of the overall
circuit implies that the in-plane torque must be an odd function
of the angle $\theta$ between $\hat{n}_{F}$ and $\hat{n}_{AF}$,
and that it can therefore be expanded in terms of a $\sin$-only
Fourier series, vanishing for both parallel and antiparallel
collinear configurations.  Most AF materials used in
magnetoelectronics are fully compensated, {\em i.e.} the
spin-density sums to zero (or nearly so) in every lattice plane perpendicular to
the current direction. In this case,
reversal of the AF moment direction is equivalent to a lateral
translation which cannot influence the current-induced torque.  It
follows that in the compensated AF case the torque is invariant
under $\theta \to \theta + \pi$, restricting its Fourier expansion
to terms proportional to $\sin(2n\theta)$.  The torque therefore
vanishes when $\hat{n}_{F}$ is perpendicular to $\hat{n}_{AF}$,
and undergoes a sign change for $\theta \to \pi - \theta$, as
illustrated in Fig. (\ref{fig:sttcartoon}).  The property that the
torque acting on a F due to a compensated AF vanishes not only for
collinear but also for perpendicular orientations is primarily
responsible for the novel current-induced torque phase diagram
that we discuss below.

\noindent {\em Current-Induced Torques for Co/NiMn}--- We employ a
non-equilibrium Green's function (NEGF) approach\cite{haneyFM} for
microscopic calculations of magneto-transport properties and
current-induced torques. Quasiparticle Hamiltonians are
constructed using density functional theory within the local spin
density approximation (extended to allow noncollinear spin
configurations), norm-conserving pseudopotentials, and an $s$,
$p$, $d$ single-zeta atomic orbital basis set.  The induced torque
per-current can be calculated atom by atom\cite{haneyFM}:
\begin{eqnarray}
\frac{{\vec {\dot S}}}{I} = \frac{\mu_B}{e} \frac {\int
dk_\parallel \sum_{\alpha,\beta}({\vec \Delta}_{\alpha,\beta}
\times {\vec m}^{tr}_{\beta,\alpha})} { \int dk_\parallel
T(\epsilon_F)}~. \label{eq:sdot}
\end{eqnarray}
The right-hand-side of Eq. (\ref{eq:sdot}) expresses the
misalignment between the non-equilibrium spin-density, $\vec
m^{tr}_{\alpha,\beta}$, and the spin-dependent part of the
exchange-correlation potential, ${\vec \Delta}_{\alpha,\beta}$.
Here $T(\epsilon_F)$ is the transmission probability,
$k_\parallel$ labels transverse channels, $\alpha, \beta$ are
orbital labels, and $\alpha$ is summed only over orbitals centered
on the atom of interest. When the many-body Hamiltonian is spin
rotationally invariant, the current-induced torque on each atom is
equal to the net spin flux out of the atom.

\begin{figure}[h!]
\begin{center}
\vskip 0.2 cm
\includegraphics[width=2.5in]{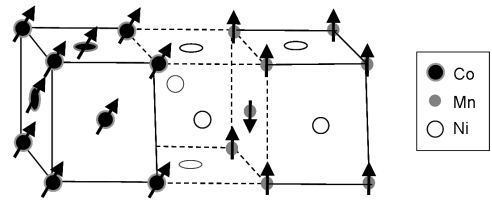}
\vskip 0.2 cm \caption{Illustration of the NiMn-Co interface
model.} \label{fig:coords}
\end{center}
\end{figure}

We apply this approach to a system with a single interface between
antiferromagnetic NiMn and ferromagnetic Co.  The crystal
structure of NiMn is face centered tetragonal, with Ni and Mn
layers alternating in the $(001)$ direction\cite{pal}. The Ni
atoms are approximately nonmagnetic, while the Mn atoms form a
compensated antiferromagnetic 2-dimensional lattice within each
plane (See Fig. (\ref{fig:coords})).  In our calculation, we use
$a=3.697~{\rm \AA}$, with a $c/a$ ratio of $0.9573$ for NiMn and,
following Ref. \cite{butlernimn}, a lattice-matched tetragonal
structure for Co with a $c/a$ ratio chosen to conserve its
experimental atomic volume. The results shown here are for current
in the (001) direction, perpendicular to the interface between Co
and Ni terminated NiMn.  The current through the interface has a
polarization $P=(T_\uparrow-T_\downarrow)
/(T_\uparrow+T_\downarrow)=6.4\%$ when the F and AF moments are
collinear, with the larger conductance for the ferromagnet
majority spins. To evaluate the current-induced torques present in
the system, we rotate the Co layer magnetization orientation by an
angle $\theta$ with respect to the NiMn moment direction and use
Eq. (\ref{eq:sdot}).

\begin{figure}[h!]
\begin{center}
\vskip 0.2 cm
\includegraphics[width=2.75in]{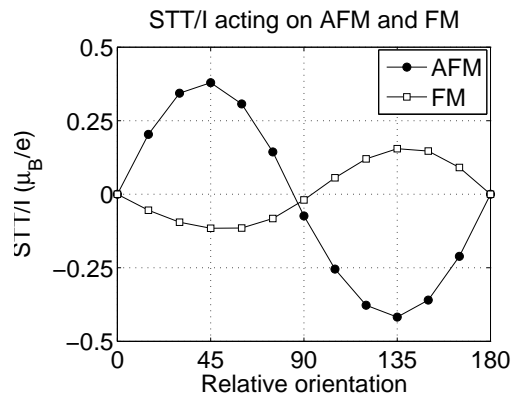}
\vskip 0.2 cm \caption{Current-induced torques per current acting
on the order parameters of the AF and F layers vs relative
orientation. In the AF the order parameter is driven by the
differences between torques on opposite sublattices.  Units are
$\mu_B/e$.} \label{fig:stt}
\end{center}
\end{figure}

Fig. (\ref{fig:stt}) shows the total torques acting on the AF and
F order parameters as a function of $\theta$.  The torque acting
on the F closely follows the form anticipated on symmetry grounds
above.  We associate the small torque at $\theta=\pi/2$ with weak
ferromagnetism which is induced in the top (Ni) layer of NiMn.  In
NiMn only the difference between the torques on the two sublattice
of the AF drives the order parameter.  The current-induced torque
tends to drive the orientation of downstream material (AF or F)
parallel with that of the upstream, and to drive the upstream
material orientation perpendicular to the downstream (so that for
electron flow from AF to F, the F tends to align to AF, and the AF
tends to become perpendicular to F, within their common plane).

We have also considered Mn terminated NiMn adjacent to Co.  In
this case the last Mn layer acquires a net magnetic moment in the
direction of Co.  The current-induced torques do not show as clean
of a $\sin 2\theta$ behavior, but a combination of $sin \theta$
and $sin 2\theta$. We conclude that the absence of odd $n$ $\sin n
\theta$ torques is closely tied to the degree of compensation at
the AF interface.

\noindent
{\em Phase Diagram for a pinned antiferromagnet}--- We now
consider the implications of this new form of current-induced
torque for systems with the usual thin film geometry, assumming
for the sake of definiteness that the AF moment direction ${\hat
n}_{AF}$ is pinned and lies in the plane, and that the external
magnetic field $H$ is applied in the same direction.  We use a
spherical coordinate system for the F moment direction, taking
${\hat n}_{AF}$ as the polar direction and the $\hat{x}$ direction
as the film normal (the demagnetizing field is denoted by $H_d$).
We assume that a non-magnetic spacer layer is placed between the F
and AF layers so that exchange
interactions negligible.  Just as in the pure F case, a spacer
layer is not expected to have a large impact on current-induced
torques. We also omit easy-axis anisotropy; its inclusion wouldn't
substantially change the picture described below. With these
ingredients the polar and azimuthal torques acting on the
ferromagnet are:
\begin{eqnarray}
\Gamma_{\theta} &=& -\frac{1}{2}\sin(\theta)\sin(2\phi)H_d +
\sin(2\theta) H_{CI}\nonumber ~;\\
\Gamma_{\phi} &=& \sin(\theta)H +
\frac{1}{2}\sin(2\theta)\cos^2(\phi) H_d~.
\end{eqnarray}
Here we have parameterized the current-induced torque by $H_{CI}$
and chosen a sign convention in which $H_{CI} <0$ when it favors
perpendicular alignment.  Steady-state solutions satisfy
$\Gamma_\phi = \Gamma_\theta=0$ and are stable for small
deviations when Gilbert damping is included. We have determined
the stability regions of the steady state solutions discussed
below by following the procedure described in Ref.
\onlinecite{BJZ2}.  We present all of our results in terms of the
dimensionless fields $h=H/H_d$ and $h_{CI}=H_{CI}/H_d$.


In the absence of the current induced torques, the magnetization
simply lines up with the magnetic field applied in the easy plane.
The influence of the current-induced torque on F is particularly
dramatic for $H_{CI} < 0$. Because the torques then tend to push
the magnetization perpendicular to ${\hat n}_{AF}$, the
field-aligned solution is stable only for:
\begin{equation}
h_{CI} \geq -\frac{\alpha}{2}\left(|h|+\frac{1}{2} \right)~
\label{eq:stability2}~,
\end{equation}
where $\alpha$ is the Gilbert damping parameter. For sufficiently
strong currents and weak external fields, a
perpendicular-to-plane steady state becomes stable:
\begin{eqnarray}
\theta &=&  \frac{\pi}{2} + h ~;\nonumber \\
\phi &=&  -2h_{CI}h  + n\pi ~.\label{eq:soln1}
\end{eqnarray}
These equations have been derived assuming that $h$ and $h_{CI}$
are small.  In the above $n$ is an even integer for solutions
which point approximately in the $+\hat x$ direction, and an odd
integer for the $-\hat x$ direction. The region of stability for
this solution is:
\begin{eqnarray}
h_{CI} \leq -\frac{\alpha}{2}  \left( \frac{h\sin h - 2\cos^2 h}{h
\sin h - \cos 2h}\right),~ \label{eq:stability1}
\end{eqnarray}
where the the fraction on the r.h.s. of the above inequality must
be negative, implying that $|h| < 0.608$, or equivalently $|H| <
\mu_0 M_s(0.608)$.

\begin{figure}[h!]
\begin{center}
\vskip 0.2 cm
\includegraphics[width=3.5in]{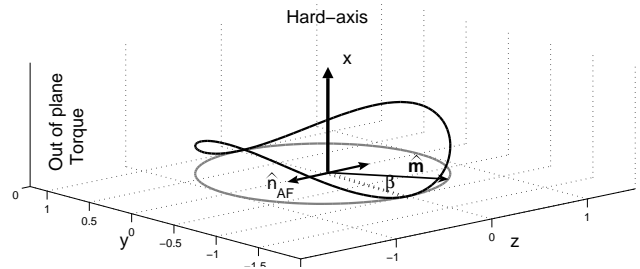}
\vskip 0.2 cm \caption{Saddle shape illustrates the out-of-plane
torque vs $\beta$ for small excursions of the magnetization
orientation $\hat m$ from the easy plane. The out-of-plane torque
is always positive. }\label{fig:hardAxis}
\end{center}
\end{figure}

The stability of this counter-intuitive stable steady state is
explained in Fig. (\ref{fig:hardAxis}).  This figure illustrates
the situation when the excursions from the easy plane are small.
For simplicity we first consider no external field.  In the
absence of the current-induced torque a small fluctuation out of
the easy plane would initiate precession about the hard axis which
damps back into the easy plane. The presence of the $\sin 2\theta$
torque, however, drives the magnetization $\hat m$ perpendicular to
$\hat{n}_{AF}$ within their common plane. As
$\hat m$ precesses around the hard-axis, this torque
vector has a component which points out of the easy plane. If the
angle between $\hat{n}_{AF}$ and the in-plane component of $\hat
m$ is $\beta$, the magnitude varies as $\Gamma_{x}=2H_{CI} m_x
\sin^2 \beta$, as shown in the figure. The crucial point is that
this torque is always positive throughout the precession. When
this torque exceeds the damping, the out-of-plane configuration is
stabilized. The eventual out-of-plane orientation can be $+\hat x$
or $-\hat x$ depending on the direction of the initial fluctuation
out of plane.  The presence of an applied field changes the
trajectory of the magnetization upon excursions from the
easy-plane.  For a sufficiently large applied field, the torque is
unable to stabilize the out-of-plane configuration, and no steady
state is reached.

\begin{figure}[h!]
\begin{center}
\vskip 0.2 cm
\includegraphics[width=3.5in]{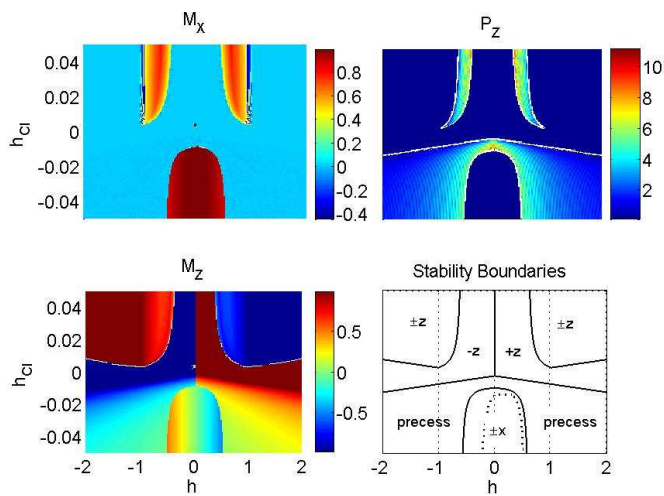}
\vskip 0.2 cm \caption{Magnetic configuration (${\rm M_x, M_z}$)
and peak of power spectrum ${\rm P_z}$ (arbitrary units) versus
applied field and current.  Also shown is stability boundaries
found analytically (the labels $\pm {\rm x}, \pm {\rm z}$ refer
also to solutions which point approximately in these directions).
The stability boundary plot also shows the reduced out-of-plane
solution space for negative to positive field sweep with a dashed
line. } \label{fig:phase}
\end{center}
\end{figure}

Interesting new steady states can in principle also be induced by the
current-induced torque for $H_{CI} > 0$.  For $|H|\leq H_d$, the steady state
stability analysis identified configurations in which the magnetization is approximately
anti-aligned with the applied field:
\begin{eqnarray}
\theta &=& \cos ^{-1}\left(-h\right)~ \nonumber \\
\phi &=& -2 h_{CI} h.\label{eq:soln3}
\end{eqnarray}
which is stable for the range of applied fields and currents:
\begin{eqnarray}
h_{CI} \geq \frac{\alpha}{2}\left( \frac{2-h^2}{3h^2-1}\right)
\label{eq:stability3}
\end{eqnarray}
For $H_{CI}>0$ and $|H|\geq H_d$, the equilibrium
solutions are $m_z = \pm 1$.  In this case the stability
condition for the magnetization anti-aligned with the field is:
\begin{eqnarray}
 h_{CI} &\geq& \frac{\alpha}{2} \left( |h| -
\frac{1 }{2}\right) ~. \label{eq:stability4}
\end{eqnarray}
These anti-aligned states occur only if the magnetization is initially nearly
anti-aligned with the applied field.  The reason for their stability
is that this form of the current-induced torque does not
distinguish between $+\hat z$ and $-\hat z$ - it merely tends to
make to direct the F to the nearest available $\hat z$-axis, even
if it's opposite to the applied field.  The region for such a
solution is shown in Fig. (\ref{fig:phase}), labelled $\pm z$.
This misaligned steady state may not be
experimentally relevant however because it occurs only when the
magnetization is initially nearly anti-aligned to an applied field
of finite magnitude $|H|>H_d\sqrt{1/3}$.

Fig. (\ref{fig:phase}) shows the $x$ and $z$ components of the
magnetization as a function of applied field and current,
determined numerically.  We have taken the damping $\alpha=.01$.
Also shown is the magnitude of the
power spectrum peak of $z(t)$ (labelled ``${\rm P_Z}$") - a
nonzero value indicates a precessing solution. Also shown is the
stability boundaries defined by Eqs. (\ref{eq:stability2},
\ref{eq:stability1}, \ref{eq:stability3}, \ref{eq:stability4}).
The numerics verify the stability of the unusual out-of-plane and
field-anti-aligned solutions.  The conversion of the dimensionless
$h_{CI}$ into a real current density for a material with
demagnetization field of 1 T is $J=(h_{CI} t) \times 3.8 \cdot
10^{9} {\rm A/cm^2}$, where $t$ is the thickness of the F layer in
nm.

The data for each $(h,~h_{CI})$ point of Fig. (\ref{fig:phase}) is
obtained beginning from an initial condition close to the solution
given by Eq. (\ref{eq:soln1}). These equilibrium solutions are not
universal attractors, and are attained for a subset of initial
conditions.  To see the effect of initial conditions, we have also
swept the applied field from negative to positive for each applied
current, using the slightly perturbed final coordinates of a
trajectory as the initial condition for the next value of applied
field.  The out-of-plane solution space is reduced, shown by the
dotted line in Fig. (\ref{fig:phase}) in the stability boundaries
plot.

We now comment on the experimental possibilities of seeing these
effects.  In the preceding analysis, we assume that the AF is
fixed.  This can be accomplished by placing a large F adjacent to
the AF, so that the AF is pinned via the exchange bias effect (the
overall stack structure would be pinning F - AF - spacer - free
F). The presence of this pinning F may influence the dynamics of
the free F, but its signature should be very distinct from the
influence of the AF layer on the free F. The orientation of the
free F should be observable from magnetoresistance effects with
the pinning ferromagnet.

A virtue of the out-of-plane F configuration is that the surface
of the AF need not be single domain for its observation. As long
as the magnetization of the AF is compensated and points in the
plane (which is the preferred direction for NiMn
\cite{sakuma,kasper}), different orientations of domains at the AF
surface should cooperatively push the F out of the plane.  The
encouraging aspect of this proposal is that the signature of the
AF current-induced torque is so unique, helping to provide a
distinguished characteristic for its observation.

\noindent {\em Acknowledgments}--- We would like to acknowledge
very helpful conversations with Maxim Tsoi and Olle Heinonen. This
work was supported in part by Seagate Corporation and by the
National Science Foundation under grant DMR-0606489, and the
computational work was supported by Texas Advanced Computational
Center (TACC).

\end{document}